\documentclass[aps,prb,twocolumn,superscriptaddress,showpacs]{revtex4}

\usepackage{color}
\usepackage{graphicx}

\begin{document}
\input epsf.sty
\title{Relationship between charge stripe order and 
structural phase transitions in La$_{1.875}$Ba$_{0.125-x}$Sr$_{x}$CuO$_{4}$}


\author{H.~Kimura}
\email[]{kimura@tagen.tohoku.ac.jp}
\affiliation{Institute of Multidisciplinary Research for Advanced Materials, 
Tohoku University, Sendai 980-8577, Japan}


\author{Y.~Noda}
\affiliation{Institute of Multidisciplinary Research for Advanced Materials, 
Tohoku University, Sendai 980-8577, Japan}

\author{H.~Goka}
\affiliation{Institute for Material Research, Tohoku University, Sendai 980-8577, Japan}

\author{M.~Fujita}
\affiliation{Institute for Material Research, Tohoku University, Sendai 980-8577, Japan}

\author{K.~Yamada}
\affiliation{Institute for Material Research, Tohoku University, Sendai 980-8577, Japan}

\author{M.~Mizumaki}
\affiliation{Japan Synchrotron Radiation Research Institute, Hyogo 679-5198, Japan}

\author{N.~Ikeda}
\affiliation{Japan Synchrotron Radiation Research Institute, Hyogo 679-5198, Japan}

\author{H.~Ohsumi}
\affiliation{Japan Synchrotron Radiation Research Institute, Hyogo 679-5198, Japan}


\date{\today}

\begin{abstract}
The nature of charge stripe order and its relationship with structural phase transitions 
were studied using synchrotron X-ray diffraction in 
La$_{1.875}$Ba$_{0.125-x}$Sr$_{x}$CuO$_{4}$ 
($0.05\leq x\leq 0.10$). 
For $x=0.05$, as temperature increased,
incommensurate superlattice peaks associated with the charge order 
disappeared just at 
the structural phase transition temperature, $T_{\rm d2}$. 
However, for $x=0.075$ 
and 0.09, 
the superlattice peaks still existed as 
a short range correlation even above $T_{\rm d2}$, 
indicating a precursor of charge ordering. 
Furthermore, temperature dependences of the superlattice peak intensity, 
correlation length, and incommensurability for 
$x=0.05$ are different from those for $x=0.075$ 
and 0.09. 
These results suggest that the transition process into the charge stripe order 
strongly correlates with the order of the structural phase transitions. 
A quantitative comparison of the structure factor associated 
with the charge order have been also made for all the samples.
\end{abstract}

\pacs{74.72.Dn, 71.45.Lr, 61.10.-i}

\maketitle

\section{Introduction}\label{intro}
For the past several years, the relationship between charge 
stripe correlations\cite{Kivelson1998} and high-$T_{\rm c}$ 
superconductivity has been intensively studied to clarify 
whether the role of the stripes for the superconductivity is positive or negative. 
Systematic studies on the La$_{1.6-x}$Nd$_{0.4}$Sr$_{x}$CuO$_{4}$ (LNSCO) system 
have shown that for the 
Low-Temperature Tetragonal (LTT; $P4_{2}/ncm$) phase, incommensurate (IC) charge- 
and magnetic orders are stabilized and compete with 
superconductivity\cite{Tranquada1995,Tranquada1996,Tranquada1997}. 
This result provided a qualitative explanation for the long-standing mystery of 
the ``1/8-problem'' in La-214 cuprates\cite{Moodenbaugh1988,Axe1989}, namely, 
the {\em ordered state} of charge stripes induced by 
the LTT transformation has a negative 
impact 
with high-$T_{\rm c}$ superconductivity. 

\begin{figure}[b]
\centerline{\epsfxsize=2.25in\epsfbox{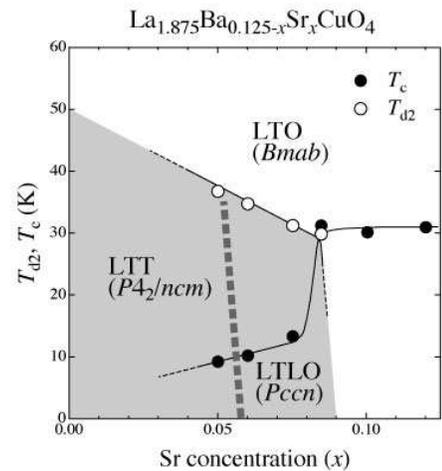}}
\caption{Structural phase transition temperature $T_{\rm d2}$ and superconducting transition temperature $T_{\rm c}$ as a function of Sr concentration for the 
La$_{1.875}$Ba$_{0.125-x}$Sr$_{x}$CuO$_{4}$ system, after 
Fujita ~{\it et al}.\cite{Fujita2002}}
\label{fig1}
\end{figure}

In the 1/8-hole-doped La$_{1.875}$Ba$_{0.125-x}$Sr$_{x}$CuO$_{4}$ (LBSCO) 
system, the crystal structure at the lowest temperature 
changes from LTT to a Low-Temperature-Orthorhombic (LTO; $Bmab$) phase via 
the Low-Temperature-Less-Orthorhombic (LTLO; $Pccn$) phase, as Sr-concentration $x$ 
increases (See Fig.~\ref{fig1}). 
Fujita~{\it et al}. have composed a detailed phase diagram of the crystal structure, 
 IC charge/magnetic order, and $T_{\rm c}$ for this system\cite{Fujita2002}, 
where the charge order is stabilized only in LTT and LTLO 
phases (gray-hatched region in Fig.~\ref{fig1}) and 
competes with superconductivity. 
On the contrary, the magnetic order in this system, 
which is robust in all the structural phases, shows 
a weak competition with the superconductivity. 

The momentum structure and the temperature evolution of the charge order in the
LBSCO system have been studied by 
Neutron diffraction\cite{Fujita2002-2} as well as X-ray diffraction\cite{Kimura2003}. 
In the LTT phase for $x=0.05$, the IC modulation wave vector ($\equiv q_{\rm ch}$) 
of the charge order is 
$(2\varepsilon, 0, 1/2)$ with High-Temperature-Tetragonal 
(HTT; $I4/mmm$) notation. 
However in the LTLO phase for $x=0.075$, 
$q_{\rm ch}$ shifts away from the tetragonal-symmetric position to 
an orthorhombic-symmetric position, giving the wave vector of 
$(+2\varepsilon, -2\eta, 1/2)$. The charge order in this system is 
stabilized just below 
the temperature where the structural phase transition from 
LTO into LTT or LTLO phase occurs $(\equiv T_{\rm d2})$. 
Further, the ordered state 
evolves as the order parameter of the LTT or LTLO structure increases. 
These facts clearly show that a strong correlation exists between the charge order 
and the crystal structure, giving rise to suppression of superconductivity. 

The charge order is detectable as lattice distortions in neutron and X-ray diffraction measurements. 
Recently, our preliminary X-ray diffraction measurements showed
that the IC superlattice peaks at $(6+2\varepsilon, 0, 11/2)$ are 
$\sim$10 times stronger in intensity than that at $(2+2\varepsilon, 0, 1/2)$. This is due to 
the amplitude of the scattering wave vector $|Q|$ and the strong $L$-dependence 
of the structure factor for the superlattice peak\cite{Kimura2003-2}, 
suggesting the importance of lattice distortions along the $c$-axis. This result indicates that 
the superlattice peak at higher-$Q$ positions is much more sensitive to the charge order 
(or the lattice distortion) than that at lower-$Q$ positions observed 
previously\cite{Kimura2003,Zimmermann1998}. This motivated us to conduct 
detailed measurements of IC superlattice peaks at 
higher-$Q$ position, especially at $(6+2\varepsilon, 0, 11/2)$ 
or $(6-2\varepsilon, 0, 17/2)$, 
for La$_{1.875}$Ba$_{0.125-x}$Sr$_{x}$CuO$_{4}$ using a synchrotron 
X-ray source for diffraction studies, which can elucidate detailed 
differences between the nature of charge 
stripes in the LTT and LTLO phases. In this paper, we show that for $x=0.075$ 
and $x=0.09$, 
short-range charge correlation starts appearing even above $T_{\rm d2}$ while the 
correlations appear just at $T_{\rm d2}$ for $x=0.05$. The results imply 
that the evolution of the charge stripes in the LTLO phase is different 
from that in the LTT phase, which relates to the order of the structural phase 
transition from the LTO to the LTT or LTLO phase. 
We also show the possibility that the displacement pattern of 
the atoms induced by the charge stripe order in the LTT phase is different from that in 
the LTLO phase. 
\section{Experimental}\label{exp}
Single crystals of LBSCO with $x=0.05$, 0.075, 0.09, and 0.10 
were cut into a cylindrical shape with dimensions of 0.43~mm diameter and 5~mm height, 
where the longest axis was parallel to the $c$-axis. 
X-ray diffraction experiments were performed at the 
Beam-line BL46XU and BL02B1\cite{Noda1998} 
of Japan Synchrotron Radiation Research Institute in SPring-8. 
The X-ray energy was tuned to 20~keV and 32.6~keV 
using a Si(111) double monochromator at BL46XU and BL02B1, respectively. 
A double platinum mirror was inserted to eliminate higher order harmonics of the X-rays. 
The samples were cooled down to 7~K using a closed-cycled $^{4}$He refrigerator. 
In this paper, the reciprocal lattice is defined in the $I{\rm 4}/mmm$ symmetry where the two 
short axes correspond to the distance between the nearest-neighbor Cu atoms along 
the in-plane Cu-O bond. 
Typical instrument resolutions along the $H$- and $K$-directions
were 0.0039~\AA$^{-1}$ and 0.0037~\AA$^{-1}$ at $Q=(6, 0, 6)$, and 
0.0038~\AA$^{-1}$ and 0.0016~\AA$^{-1}$ at $Q=(4, 0, 0)$, respectively. 
In the present study, we obtained nearly single-domain orthorhombic crystals for 
$x=0.075$, 0.09, and 0.10. Note that the measurements for $x=0.05$ and 0.075 
were done at BL46XU and those for $x=0.09$ and 0.10 were carried out at BL02B1.

As mentioned in Sec.~\ref{intro}, we focused on the measurements of the superlattice peaks at 
$Q_{\rm ch}=(h\pm2\varepsilon, 0, l/2)$ with $h=6, 8$ and $l=11, 17$ 
in the present study. 
$(5, 0, 0)$ and $(7, 0, 0)$ Bragg reflections, 
which appears only in the LTT and LTLO phases and corresponds to the order parameter for these 
phases, were also measured to compare the phase transition 
of the charge order with that of the crystal structure. 
Note that we obtained a much better signal-to-noise ratio than that 
in the previous study\cite{Kimura2003} 
by measuring the superlattice peaks at $L=11/2, 17/2$. Thus in this paper, we show $q$-profiles 
as a raw data, not as a subtracted data. 
\section{Results}\label{res}
\subsection{$Q$-dependence}\label{Qdep}
\begin{figure}[b]
\centerline{\epsfxsize=2.45in\epsfbox{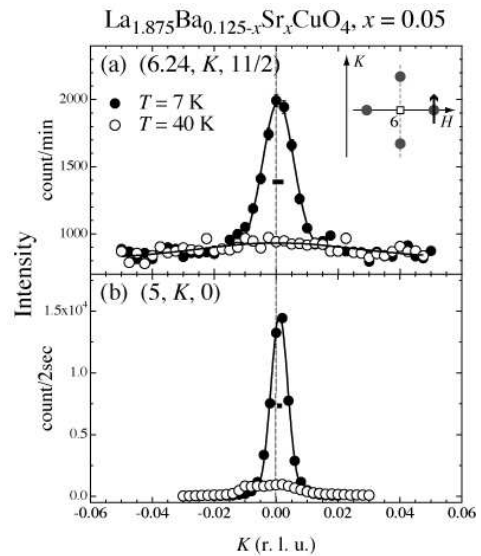}}
\caption{$q$-profiles along the $K$-direction of (a); superlattice peak through 
$Q_{\rm ch}=(6.24, 0, 11/2)$, (b); $(5, 0, 0)$ Bragg reflection 
for $x=0.05$. Scan trajectory and confirmed peak positions of superlattice peaks are 
illustrated in the inset of (a). Closed- and open circles correspond to the data taken at 
7~K and 40~K, respectively. 
Bold horizontal lines correspond to the instrument resolutions.}
\label{fig2}
\end{figure}

$Q$-scan profiles along the $K$-direction of the superlattice peak and the $(5, 0, 0)$ peak for 
$x=0.05$, taken at $T=7$~K and 40~K, 
are shown in Fig.~\ref{fig2}. 
The trajectory of the $q$-scan for the superlattice peak 
is shown in the inset of Fig.~\ref{fig2}(a). $H$- and $K$-scans for 
the superlattice peak at $T=7$~K confirmed that a quartet of 
superlattice peaks are located exactly at 
$Q_{\rm ch}=(6\pm2\varepsilon, 0, L/2)$, $(6, \pm2\varepsilon, L/2)$ with 
$2\varepsilon=0.2390(5)$~r.l.u., for which the geometry is 
consistent with the crystal symmetry of the LTT structure. 
\begin{figure}[t]
\centerline{\epsfxsize=2.45in\epsfbox{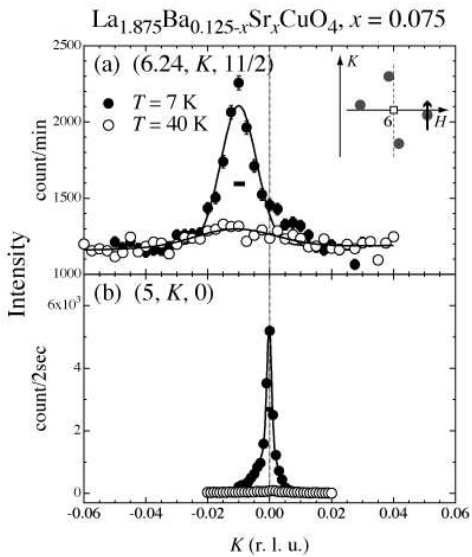}}
\caption{$q$-profiles along the $K$-direction of (a); superlattice peak through 
$Q_{\rm ch}=(6.24, -0.01, 11/2)$, (b); $(5, 0, 0)$ Bragg reflection for $x=0.075$. 
Scan trajectory and confirmed peak positions of superlattice peaks are 
illustrated in the inset of (a). 
Closed- and open circles correspond to the data taken at 7~K and 40~K, respectively. 
Bold horizontal lines 
correspond to the instrument resolutions.}
\label{fig3}
\end{figure}
The observed line-width along the $K$-direction for the superlattice peak 
is apparently broader than the instrument resolution (denoted 
in the figure as a bold horizontal line), giving a finite correlation 
length for the charge correlations. Note that the line-width along the $H$-direction for 
the superlattice peaks becomes also broader. 
\begin{figure}[t]
\centerline{\epsfxsize=2.45in\epsfbox{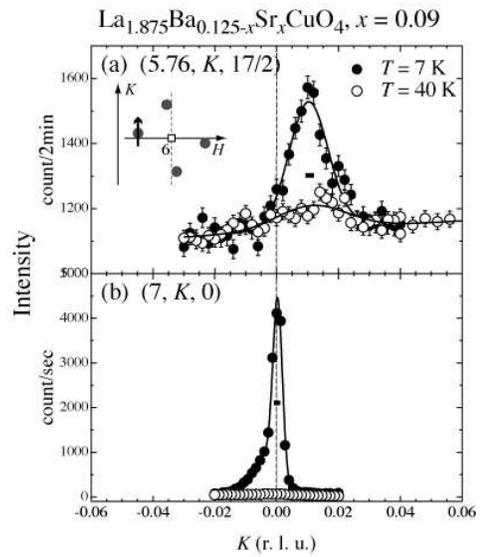}}
\caption{$q$-profiles along the $K$-direction of (a); superlattice peak through 
$Q_{\rm ch}=(5.76, 0.01, 17/2)$, (b); $(7, 0, 0)$ Bragg reflection for $x=0.09$. 
Scan trajectory and confirmed peak positions of superlattice peaks are 
illustrated in the inset of (a). 
Closed- and open circles correspond to the data taken at 7~K and 40~K, respectively. 
Bold horizontal lines correspond to the instrument resolutions.}
\label{fig4}
\end{figure}
As a result, the correlation lengths of the charge 
order along $a$- and $b$-axis ($\equiv\xi_{\rm ch}(a)$, $\xi_{\rm ch}(b)$) are 
98$\pm$4~\AA\ and 110$\pm$4~\AA\ at $T=7$~K, respectively. 
For the $(5, 0, 0)$ peak, the line-width along the $K$-direction 
is broader than the instrumental resolution while the line-width along $H$-direction 
reaches the resolution limit. Thus the correlation length for the LTT structure, 
$\xi_{a}$ and $\xi_{b}$, are estimated to be $>300$~\AA\ and 196$\pm$5~\AA, 
respectively, indicating a large anisotropy of the structural coherence or 
a mosaic spread due to a local disorder at the LTT phase. 
At $T=40$~K, just below $T_{\rm d2}$, both the superlattice peak 
and the $(5, 0, 0)$ peak almost vanish, indicating that 
the charge order appears when the structural phase transition into the LTT phase 
occurs. 

Figures~\ref{fig3}(a) and (b) show $q$-scan profiles along the $K$-direction of 
the superlattice peak and the $(5, 0, 0)$ peak for $x=0.075$, respectively, also 
taken at $T=7$~K and 40~K. 
The trajectory of the $q$-scan for the superlattice peak 
is displayed in the inset of Fig.~\ref{fig3}(a). Since the 
single-domain-LTLO phase was obtained for the $x=0.075$ sample, we confirmed that 
a shift of the superlattice peaks from the highly symmetric 
axis clearly exists and the exact peak position is determined as 
$Q_{\rm ch}=(6\pm2\varepsilon, \mp2\eta, L/2)$, 
$(6\mp2\eta, \pm2\varepsilon, L/2)$ with 
$2\varepsilon=0.2360(5)$~r.l.u. and $2\eta=0.0100(5)$~r.l.u. 
The observed line-width along the $K$-direction  
for the superlattice peak is much broader than the resolution, of 
which value is almost comparable 
to that for $x=0.05$. On the other hand, the line-width for the 
$(5, 0, 0)$ peak is resolution-limited, which is much sharper 
than that for $x=0.05$. Therefore, 
$\xi_{\rm ch}(a)$ and $\xi_{\rm ch}(b)$ for the charge order are 
104$\pm$5~\AA\ and 100$\pm$7~\AA, respectively, while $\xi_{a}$ and $\xi_{b}$ for 
the LTLO structural coherence become long-ranged, which is in contrast with the 
results for $x=0.05$. 
At $T=40$~K, far above $T_{\rm d2}$, the broad superlattice peak 
clearly remains while the $(5, 0, 0)$ peak disappears, suggesting that the charge order 
exists even above $T_{\rm d2}$ with a short range correlation. 

Figures~\ref{fig4}(a) and (b) show $q$-scan profiles at 
$T=7$~K and 40~K along the $K$-direction of 
the superlattice peak through $(5.76, 0.01, 17/2)$ and the $(7, 0, 0)$ peak for $x=0.09$, 
respectively, taken at BL02B1. The trajectory of the $q$-scan for the superlattice peak 
is displayed in the inset of Fig.~\ref{fig4}(a). 
This sample also had the single domained structure at LTLO phase. 
Thus the exact values of $2\varepsilon$ and $2\eta$ are obtained as 
$0.2403(5)$ and $0.0103(3)$, respectively. As seen in Fig.~\ref{fig4}(a), the line-width 
along $K$-direction 
is much broader than the resolution, which is also seen in the line width along $H$. 
As a result, $\xi_{\rm ch}(a)$ and $\xi_{\rm ch}(b)$ for the charge order become 
80$\pm$4~\AA\ and 80$\pm$5~\AA, respectively, which is shorter than those for 
$x=0.05$ and $x=0.075$. As for the $(7, 0, 0)$ peak, the line-width is somewhat broader than 
the resolution but $\xi_{a}$ and $\xi_{b}$ still extend over 200~\AA. 
Although the $(5, 0, 0)$ peak completely disappears at $T=40$~K, 
the broad superlattice peak still clearly exists, which is consistent with the results of $x=0.075$. 
We had observed no superlattice peak in the $x=0.10$ sample but observed quite weak 
(5, 0, 0) peak, indicating that the development of the order parameter for the LTLO phase is 
too small to stabilize the charge order. 

In our previous paper, we argued for the anisotropy of $\xi_{\rm ch}(a)$ and 
$\xi_{\rm ch}(b)$, based on the comparison with the $\xi_{a}$ and $\xi_{b}$ 
of LTT/LTLO structure\cite{Kimura2003}. However, the present study, 
under the fine resolution in $q$-space, has shown that 
the structural coherence for the LTT phase is apparently different from that for the LTLO phase, 
which was not observed in the previous experiment. Therefore in the present study, we 
evaluated the value of $\xi_{\rm ch}(a)$ and $\xi_{\rm ch}(b)$ by 
comparing the observed line-widths 
of fundamental Bragg peaks taken at room temperature, which corresponds to 
the accurate instrument resolutions. 
\subsection{$T$-dependence}\label{Tdep}
The temperature dependence of integrated intensity, line-width, 
$2\varepsilon$, and $2\eta$
were measured in detail for the IC superlattice peaks for $x=0.05$, $x=0.075$, 
and $x=0.09$. 
For the $(5, 0, 0)$ and $(7, 0, 0)$ peak, the temperature dependence of 
integrated intensity and line-width were measured. 
All the measurements were performed during heating process. 
\begin{figure}[t]
\centerline{\epsfxsize=2.45in\epsfbox{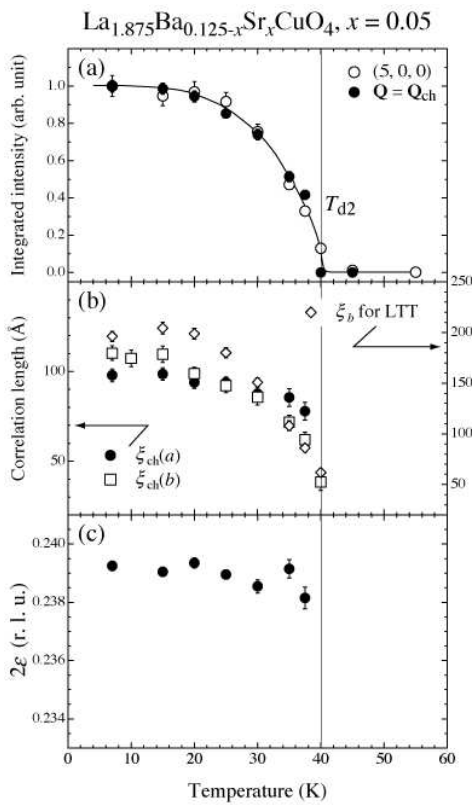}}
\caption{Temperature dependences of (a); integrated intensity for 
the superlattice peak (closed circles) and the (5, 0, 0) peak (open circles), 
(b); correlation length along the $a$-axis (closed circles), 
$b$-axis (open squares), (c); 2$\varepsilon$ for $x=0.05$. 
The correlation length along $b$-axis for LTT structure 
is plotted in (b) with open diamonds against a right vertical axis. 
The solid curve in (a) is to guide the eye.}
\label{fig5}
\end{figure}
The results for $x=0.05$ are summarized in Fig.~\ref{fig5}. 
Figure~\ref{fig5}(a) shows the temperature dependence of 
integrated intensity for the superlattice peak at 
$Q_{\rm ch}=(6.239, 0, 11/2)$ and 
the $(5, 0, 0)$ peak, where the intensities are normalized at 7~K. 
It is seen that the evolution of the intensity for 
the superlattice peak agrees well with that for the $(5, 0, 0)$ peak, 
apparently indicating that 
the charge order 
appears just at 
$T_{\rm d2}$ ($\sim40$~K) 
and the order parameters for the charge order and 
the LTT structure are strongly associated with each other. 
$\xi_{\rm ch}(a)$ and $\xi_{\rm ch}(b)$ for the charge order and $\xi_{b}$ for the LTT structure 
as a function of temperature are plotted in Fig.~\ref{fig5}(b), 
for which values are obtained from the inverse of the intrinsic line-width. 
Note that $\xi_{a}$ for the LTT 
phase 
cannot be plotted 
in the figure because the correlation along the $a$-axis becomes almost a long-range one below $T_{\rm d2}$. 
As temperature decreases, both $\xi_{\rm ch}(a)$ and $\xi_{\rm ch}(b)$
increase and show a nearly isotropic correlation 
with the length of $\sim100$\ \AA. 
In the case of $\xi_{\rm ch}(b)$, the temperature variation is 
quite similar to the development of $\xi_{b}$ for the LTT structure, implying 
that the growth of the charge correlation follows the evolution of 
the LTT structural coherence along the $b$-axis. 
As seen in Fig.~\ref{fig5}(c), the incommensurability $2\varepsilon$ for $x=0.05$ 
is nearly constant for all temperature regions below $T_{\rm d2}$. 

\begin{figure}[t]
\centerline{\epsfxsize=2.45in\epsfbox{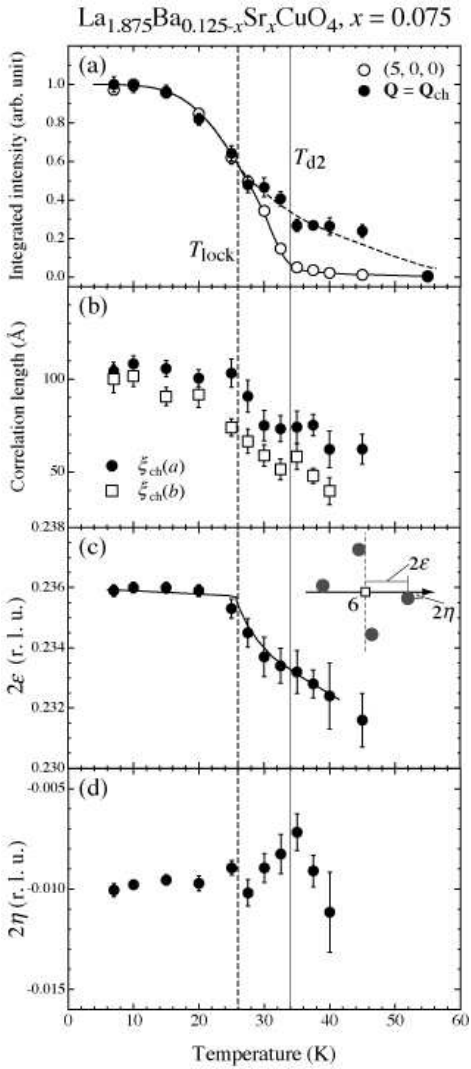}}
\caption{Temperature dependences of (a); integrated intensity for 
the superlattice peak (closed circles) and the (5, 0, 0) peak (open circles), 
(b); correlation length along the $a$-axis (closed circles) and 
$b$-axis (open squares), (c); 2$\varepsilon$, (d); 2$\eta$ 
for $x=0.075$. Definitions of 2$\varepsilon$ and 2$\eta$ are 
shown in the inset of (c). 
The bold- and dashed curves are guides to the eye.}
\label{fig6}
\end{figure}

Figure~\ref{fig6} shows the summary of results for $x=0.075$. 
The integrated intensity of the superlattice peak and the $(5, 0, 0)$ peak 
are depicted in Fig.~\ref{fig6}(a) as a function of temperature. 
The $(5, 0, 0)$ peak starts growing below $T_{\rm d2}$ ($\sim34$~K) 
where the structural phase transition 
from the LTO to the LTLO phase occurs, while the superlattice peak appears 
at a much higher temperature than $T_{\rm d2}$. 
In the lower temperature region, the temperature dependence of the superlattice peak intensity 
coincides with that for the $(5, 0, 0)$ peak intensity, which is also seen 
in the results for $x=0.05$. 
However, above $T\sim26$~K (indicated in Fig.~\ref{fig6} as a vertical dashed line), 
the superlattice peak intensity decreases more gradually than the decay of 
\begin{figure}[t]
\centerline{\epsfxsize=2.45in\epsfbox{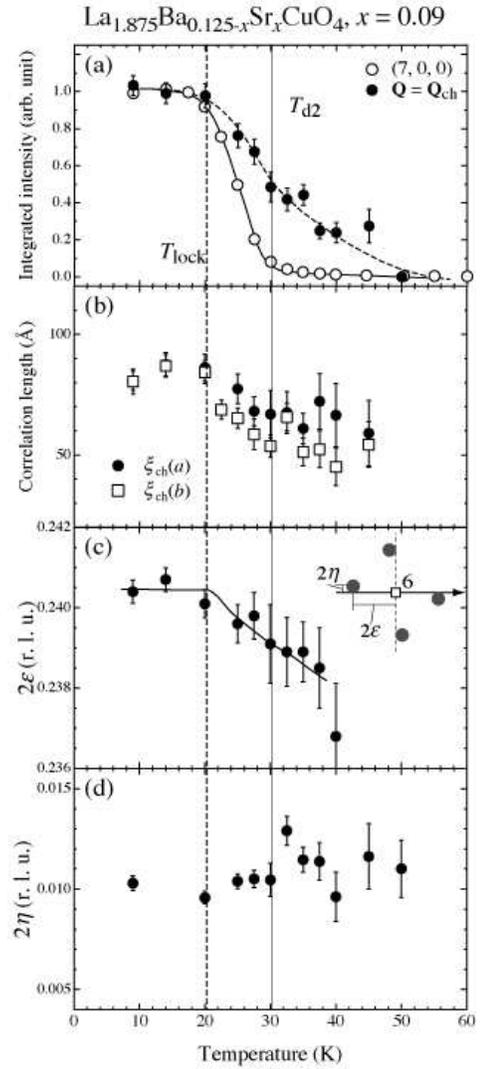}}
\caption{Temperature dependences of (a); integrated intensity for 
the superlattice peak (closed circles) and the (7, 0, 0) peak (open circles), 
(b); correlation length along the $a$-axis (closed circles) and 
$b$-axis (open squares), (c); 2$\varepsilon$, (d); 2$\eta$ 
for $x=0.09$. Definitions of 2$\varepsilon$ and 2$\eta$ are 
shown in the inset of (c). 
The bold- and dashed curves are guides to the eye.}
\label{fig7}
\end{figure}
the $(5, 0, 0)$ peak with increasing temperature. 
The temperature dependence of the correlation length for the charge order 
is plotted in Fig.~\ref{fig6}(b). Both $\xi_{a}$ and $\xi_{b}$ for LTLO structural 
coherence are not shown because the correlations along $a$- and $b$-axis reach 
at least 300~\AA\ for all temperature regions below $T_{\rm d2}$. 
At the lowest temperature, 
the correlation of the charge order is nearly isotropic with the length of 
$\sim100$~\AA\ which is almost identical to the charge correlation for $x=0.05$.
However, one can seen in Fig.~\ref{fig6}(b) that 
the correlation length suddenly changes around $T\sim26$~K, 
which is not seen in the charge correlation for $x=0.05$. 
As shown in Figs.~\ref{fig6}(c) and (d), the incommensurability $2\varepsilon$ 
starts increasing with decreasing temperature and saturates below $\sim26$~K
while the peak shift $2\eta$ from 
the fundamental axis is almost temperature independent. These results imply that 
the charge order initially appears as short range correlations 
well above $T_{\rm d2}$ and the correlation starts 
extending well below $T_{\rm d2}$, where the IC modulation vector for 
the charge order is locked into $2\varepsilon=0.236$~r.l.u. 
In this paper, we defined the temperature where the $Q_{\rm ch}$ 
is locked as $T_{\rm lock}$. 

The summary of the results for $x=0.09$ is shown in Fig.~\ref{fig7}. 
The temperature dependence of the integrated intensity for the superlattice peak and 
the $(7, 0, 0)$ peak are displayed in Fig.~\ref{fig7}(a). The intensities are normalized 
by the values taken at $T=7$~K. As temperature decreases, 
the structure phase transition into LTLO phase occurs at $T_{\rm d2}$ ($\sim 30$~K) 
which follows the appearance of the superlattice peak. Around the lowest temperature, 
the temperature evolution of the superlattice peak is almost coincides 
with that of the $(7, 0, 0)$ peak. However, above $T\sim20$~K, denoted by the dashed line in 
the figure, the temperature dependence 
of the superlattice peak is considerably different from that of the $(7, 0, 0)$ peak. 
As seen in Fig.~\ref{fig7}(b), a characteristic change also occurs in 
the temperature dependence of $\xi_{\rm ch}(a)$ and $\xi_{\rm ch}(b)$, where 
the correlation length suddenly extends. Furthermore, the 
incommensurability $2\varepsilon$ saturates into 0.24 below 20~K (See Fig.~\ref{fig7}(c)). 
These behaviors show that there is a characteristic temperature $T_{\rm lock}$ also in $x=0.09$, 
which is lower than that in $x=0.075$. 
At the lowest temperature, $\xi_{\rm ch}$ becomes almost isotropic but the correlation length 
remain $\sim 80$~\AA, which is shorter than that in both $x=0.05$ and $x=0.075$. 
The result implies that the order parameter of the charge order for $x=0.09$ is reduced 
comparing with that for $x=0.05$ and $x=0.075$. 
As shown in Fig.~\ref{fig7}(d), $2\eta$ is also temperature independent. 
\section{Discussion and Conclusions}\label{dis}
\subsection{Modulation wave vector of a charge order}\label{dis1}
We first refer to the IC modulation wave vectors of the charge order. 
The present study confirmed that 
the modulation vector $q_{\rm ch}$ for $x=0.05$, $x=0.075$, and $x=0.09$ is 
$(0.239, 0, 1/2)$, $(0.236, -0.010, 1/2)$, and $(0.240, -0.010, 1/2)$, 
respectively\cite{Kimura2003-3}. 
Note that the concentration of (Ba + Sr) ions for the $x=0.075$ sample is 
roughly estimated to be 0.117 by ICP emission spectroscopy, which is nearly consistent with the 
$\varepsilon (=0.118)$ for $x=0.075$. Therefore, the effective concentration of doped-holes 
almost coincides with the incommensurability of the modulation wave vector, 
which suggests a 1/4-filling configuration in the charge stripes. 

$q_{\rm ch}$ for $x=0.075$ 
and $x=0.09$ 
shows that the IC modulation wave vector does not lie 
on the fundamental reciprocal axis (i. e. $H$, or $K$-axis), which has been originally 
found in the IC magnetic order of La$_{2}$CuO$_{4+y}$\cite{Lee1999}. 
This shift from the symmetry axis is quantified by the angle of $\theta_{\rm Y}$ 
between the modulation wave vector and the $H$ (or $K)$-axis. The definition of 
\begin{figure}[t]
\centerline{\epsfxsize=2.45in\epsfbox{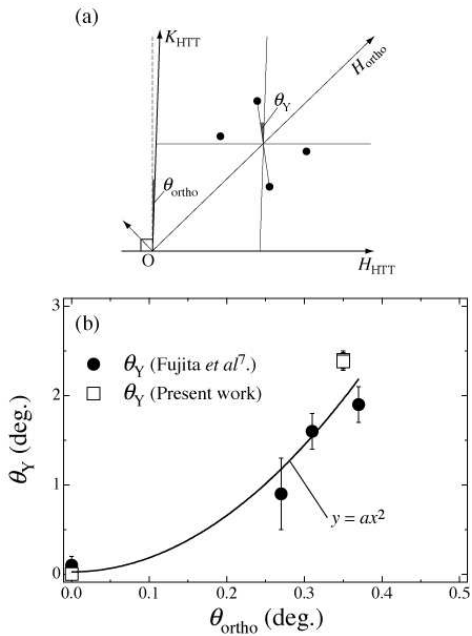}}
\caption{(a); Schematic representation of the geometry 
of the IC magnetic peaks and the definitions of $\theta_{\rm ortho}$ and 
$\theta_{\rm Y}$. (b); $\theta_{\rm Y}$ as a function of 
$\theta_{\rm ortho}$. Closed circles and open squares were obtained by 
Fujita {\it et al}\cite{Fujita2002}. and from the present study, respectively.}
\label{fig8}
\end{figure}
$\theta_{\rm Y}$ is displayed in Fig.~\ref{fig8}(a). 
Fujita {\it et al}. have found\cite{Fujita2002-2} that the amplitude of 
$\theta_{\rm Y}$ in the LBSCO system is proportional to the square value 
of the orthorhombic distortion ($\equiv \theta_{\rm ortho}$), which is quantified as 
the deviation from 90$^{\circ}$ in the angle 
between $H$- and $K$-axis in the HTT unit (See Fig.~\ref{fig8}(a)). 
As shown in Fig.~\ref{fig8}(b), $\theta_{\rm Y}$ as a 
function of $\theta_{\rm ortho}$ 
obtained by Fujita {\it et al}. (closed circles) agrees well with 
the results obtained in the present studys (open squares). 
Note that $\theta_{\rm Y}$ and $\theta_{\rm ortho}$ for 
$x=0.075$ and $x=0.09$ are almost coincide within the experimental error. 
Theoretical work 
based on {\em fermiology} has pointed out that $\theta_{\rm Y}$ can be understood as 
an anisotropy of the second nearest-neighbor transfer integral due to the orthorhombic symmetry in 
the CuO$_{2}$ plane\cite{Yamase2000-1}. 
However, a detailed displacement pattern of oxygen atoms 
associated with the charge order should be resolved to explain the origin of the peak shift. 
\subsection{Order parameter of structural phase transitions and a charge order}\label{dis2}
Structural phase transitions from the LTO to LTLO, and from the 
LTO to LTT phase, in La-214 cuprates can be understood in terms of the Landau-Ginzburg 
free energy of the order parameter, which is described by the amplitude of the tilting of 
CuO$_{6}$ octahedra\cite{Axe1989,Ting1990}. In this framework, the LTO-LTT transition 
shows a first-order phase transition while the LTO-LTLO transition 
should be a second-order phase transition, which depends on the sign of the eighth-order term 
in expanding the Landau free energy. 
Therefore, the structural phase transition in $x=0.05$ is a first-order while 
$x=0.075$ and $x=0.09$ should show a second-order phase transition. 
X-ray powder diffraction analyses have shown that 
the LTO-LTT transition in La$_{1.875}$Ba$_{0.125}$CuO$_{4}$ is a first order transition, where 
both the LTO and LTT phases coexist and the volume fraction of the LTT phase 
increases with decreasing temperature\cite{Axe1989,Billinge1993}. Therefore, 
the temperature dependence of (5, 0, 0) intensity and $\xi_{b}$ for $x=0.05$, shown in 
Fig.~\ref{fig5}(a) and (b), can be regarded as the change of the 
volume fraction of LTO and LTT structure. 
Based on this argument, it is plausible that the 
difference between the temperature evolution near $T_{\rm d2}$ of the charge order 
for $x=0.05$ and that for $x=0.075$ 
and $x=0.09$ 
closely correlates with the 
order of each structural phase transition; In the case of $x=0.05$, 
there is no critical phenomenon associated with the charge order because the 
structural phase transition is a first order one. On the contrary, for $x=0.075$ 
and $x=0.09$, 
the short-range charge correlation above $T_{\rm d2}$ is induced by the 
successive increase in structural instabilities or fluctuations near 
the second-order LTO-LTLO phase transition. 
It should be noted that X-ray diffraction integrates over 
both elastic and inelastic scattering. Therefore there is also a possibility that 
the weak signals above $T_{\rm d2}$ indicate dynamical charge (stripe) correlations. 
\subsection{Correlation length}\label{dis3}
The coherence of the LTT structure for $x=0.05$ along the $b$-axis ($\xi_{b}$) 
extends with decreasing temperature but remains within a finite length ($\sim200$~\AA). 
In contrast, the coherence of the LTLO structure for $x=0.075$ 
and $x=0.09$ is almost long-ranged. 
The correlation length of the charge order, however, is 
less than $\sim$100~\AA\ for all the samples, 
which is much shorter 
than the structural coherence. These results show that the charge stripes in this system 
are essentially glassy or topologically disturbed. Comparing $\xi_{\rm ch}$ with 
the correlation length of the magnetic order 
($\equiv\xi_{\rm spin}$) obtained by the previous neutron scattering 
study\cite{Fujita2002-2}, we thus obtain the ratio; $\xi_{\rm spin}/\xi_{\rm ch}> 2$. 
Note that in LNSCO\cite{Tranquada1999} and 
La$_{5/3}$Sr$_{1/3}$NiO$_{4}$\cite{Lee2001}, $\xi_{\rm spin}/\xi_{\rm ch}$ is about 
4 and 3, respectively. Zachar {\it et al}. have argued, from a theoretical standpoint,
that in the case of $1<\xi_{\rm spin}/\xi_{\rm ch}\lesssim 4$, 
charge stripes are disordered by non-topological elastic deformations, 
resulting in a Bragg-glass-like state or a discommensuration\cite{Zachar2000}. 

Charge correlation $\xi_{\rm ch}$ for $x=0.075$ and $x=0.09$ becomes longer 
below around $T_{\rm lock}$, where the evolution of the superlattice peak is 
superposed with that of $(5, 0, 0)$/$(7, 0, 0)$ peaks and the IC modulation wave vector 
is locked. 
Based on the stripe model, 
$\xi_{\rm ch}(a)$ denotes the deformation of the periodicity or 
the discommensuration for charge stripes and 
$\xi_{\rm ch}(b)$ corresponds to the mosaicity of stripes. 
From this point of view, the results for $x=0.075$ 
and $x=0.09$ 
indicate that the deformation of the stripe-periodicity 
and the stripe-mosaicity 
are reduced as temperature 
decreases and 2$\varepsilon$ is pinned finally at the value of 
hole concentration. If the 1/4-filling is robust in the charge stripes, the temperature 
variation of 2$\varepsilon$ indicates that the number of localized holes 
increases with decreasing temperature, which immobilizes charge stripes. 
The locking of the incommensurability is also seen 
in LNSCO\cite{Zimmermann1998} and striped nickelates\cite{Kajimoto2000}. 
However, the connection between the locking effect and the structural phase transition 
was not observed in either case. 
Note that the temperature dependence of the 
incommensurability for magnetic order should be compared with 
that of 2$\varepsilon$ in the $x=0.075$ 
and $x=0.09$ 
samples to clarify the microscopic 
interrelation between the spin- and charge correlations. 
\subsection{Comparison of structure factors}\label{F}
We finally compare quantitatively the structure factors of the lattice distortion 
associated with the charge order for $0.05\leq x\leq 0.10$. 
\begin{figure}[t]
\centerline{\epsfxsize=2.45in\epsfbox{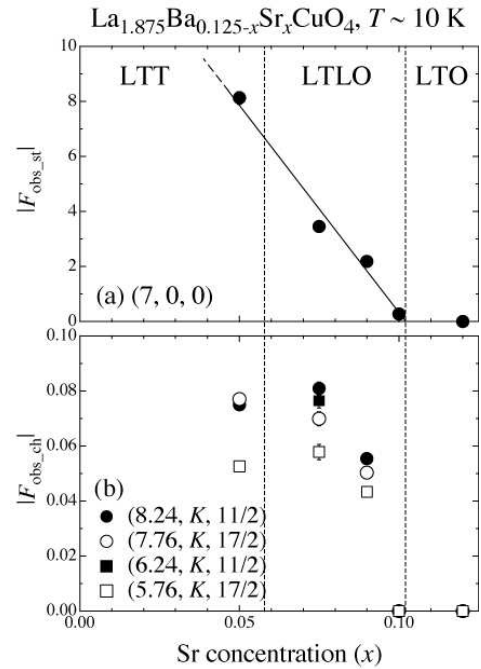}}
\caption{Absolute value of structure factor at $\sim 10$~K. 
for (a); (7, 0, 0) peaks ($\equiv|F_{\rm obs\_st}|$) 
and (b); IC superlattice peaks ($\equiv|F_{\rm obs\_ch}|$) taken at 
four reciprocal lattice points as a function of Sr concentration. 
The results of Ba-free $x=0.12$ (LSCO $x=0.12$) 
are also shown in the figures\cite{Kimura2004}.}
\label{fig9}
\end{figure}
The integrated intensity of the superlattice peaks 
were converted into the absolute value of the structure factor $|F_{\rm obs\_ch}|$ 
using the scale factor obtained from the measurements of fundamental Bragg intensities. 
The absolute value of the structure factor for the LTT/LTLO 
structure ($\equiv|F_{\rm obs\_st}|$) were also obtained to compare 
with each $|F_{\rm obs\_ch}|$. Figure~\ref{fig9} shows $|F_{\rm obs\_st}|$ and 
$|F_{\rm obs\_ch}|$ as a function of Sr concentration. 
The figure includes the result of 
Ba-free $x=0.12$ (LSCO $x=0.12$) taken previously\cite{Kimura2004}. 
As seen in Fig.~\ref{fig9}(a), $|F_{\rm obs\_st}|$ linearly increases with decreasing 
Sr concentration. It shows that atomic displacements of La (Ba, Sr) and O associated with 
the LTT/LTLO structure increase as Sr concentration decreases. $|F_{\rm obs\_ch}|$ also shows 
the linear relation with Sr concentration in the LTLO phase. 
Thus we speculate that the charge order in the LTLO phase 
becomes more stable as a pinning potential 
in the CuO$_{2}$ plane increases, which is consistent with the fact that $T_{\rm lock}$ becomes 
higher as Sr concentration increases. However, $|F_{\rm obs\_ch}|$ of $x=0.05$ in the LTT phase 
is comparable with that of $x=0.075$ in the LTLO phase while $|F_{\rm obs\_st}|$ of 
$x=0.05$ is much stronger than that of $x=0.075$. The result implies that the structure factor 
of the lattice distortion associated with the charge order in the LTT structure is different with that 
in the LTLO phase; namely, the displacement pattern of oxygen atoms in the LTT phase is 
different from that in the LTLO phase. 
\subsection{Conclusions}\label{conc}
The relationship between charge stripes and structural phase transitions 
was systematically studied for 
La$_{1.875}$Ba$_{0.125-x}$Sr$_{x}$CuO$_{4}$ 
with $0.05\leq x\leq 0.10$. 
We have found that 
the short range charge correlations appear above $T_{\rm d2}$ for $x=0.075$ 
and $x=0.09$ while 
the correlations starts growing just at $T_{\rm d2}$ for $x=0.05$. 
Furthermore in both the $x=0.075$ 
and $x=0.09$ 
samples, the temperature dependence of the correlation length and 
the incommensurability are different from 
those for the $x=0.05$ sample. These facts are closely related with the order of 
the structural phase transitions from the LTO phase to the LTLO or LTT phases. 
The quantitative comparison of the structure factors for the charge order and 
the LTT/LTLO structure reveals that the charge order becomes more robust as 
the order parameter of the LTLO structure increases. Comparison of $|F_{\rm obs\_ch}|$ for 
tetragonal $x=0.05$ with that for orthorhombic $x=0.075$ indicates that 
the displacement pattern induced by the charge order in the LTT phase is different from that 
in the LTLO phase. A detailed structure analysis in the charge ordered phase 
is required to discuss more quantitatively. The structure analysis for $x=0.05$ 
is now in progress. Thus the detailed displacement pattern induced by the charge order 
will be clarified in the near future.
\begin{acknowledgments}
We thank M. Ito, K. Machida, H. Yamase, and O. Zacher for valuable discussions.
This work was supported in part by a Grant-In-Aid for Young 
Scientists B (13740216 and 15740194), 
Scientific research B (14340105), 
Scientific Research on Priority Areas (12046239), and 
Creative Scientific Research (13NP0201) from the Japanese Ministry of Education, 
Science, Sports and Culture, and by the Core Research for Evolutional Science and 
Technology (CREST) from the Japan Science and Technology Corporation.
The synchrotron X-ray experiments were carried out at the SPring-8 facility 
(Proposal No. 2002A0314, R03A46XU, 2003B0117, and 
2004A2117). 
\end{acknowledgments}


\begin{references}
\bibitem{Kimura2003-3}The $q_{\rm ch}$ for 
$x=0.075$ has been obtained as $(0.24\ -0.007\ 1/2)$ at the 
previous experiment\cite{Kimura2003}. 
However, the result of the present study 
is more reliable because the instrument resolutions and the statistics are much 
more improved than those of the previous work.
\end{references}

\end{document}